\documentclass[11pt]{article}

\usepackage{bm}
\usepackage[normalem]{ulem}
\usepackage[utf8]{inputenc}
\usepackage[T1]{fontenc}
\usepackage{mathptmx}
\usepackage{etoolbox}
\usepackage{xcolor}
\usepackage{physics}

\usepackage[a4paper, margin=0.7in]{geometry}
\usepackage{authblk}
\usepackage{graphicx}
\usepackage{dcolumn}
\usepackage{amsmath}
\usepackage{amssymb}
\usepackage{textcomp}
\usepackage{verbatim}
\usepackage{float}
\usepackage[symbol]{footmisc}
\setfnsymbol{wiley}
\usepackage{hyperref}
\hypersetup{
    colorlinks=true,
    linkcolor=blue,
    filecolor=blue,
    urlcolor=blue,
    citecolor = blue,
    pdfauthor=author,
}

\usepackage{multirow}
\usepackage[T1]{fontenc}
\usepackage[normalem]{ulem}
\usepackage[numbers,sort&compress]{natbib}

\begin{document}

\title{Topological Nanophononic Interface States Using High-order Bandgaps in the One-Dimensional Su-Schrieffer-Heeger Model}%

\author[1]{A. Rodriguez}
\author[1]{K. Papatryfonos}
\author[1]{E.R. Cardozo de Oliveira}
\author[1]{N. D. Lanzillotti-Kimura\footnote{daniel.kimura@c2n.upsaclay.fr}}

\affil[1]{Université Paris-Saclay, CNRS, Centre de Nanosciences et de Nanotechnologies, 91120 Palaiseau, France}

\date{}
\maketitle
\begin{abstract}

Topological interface states in periodic lattices have emerged as valuable assets in the fields of electronics, photonics, and phononics, owing to their inherent robustness against disorder. Unlike electronics and photonics, the linear dispersion relation of hypersound offers an ideal framework for investigating higher-order bandgaps. In this work, we propose a design strategy for the generation and manipulation of topological nanophononic interface states within high-order bandgaps of GaAs/AlAs multilayered structures. These states arise from the band inversion of two concatenated superlattices that exhibit inverted spatial mode symmetries around the bandgap. By adjusting the thickness ratio of the unit cells in these superlattices, we are able to engineer interface states in different bandgaps, enabling the development of versatile topological devices spanning a wide frequency range. Moreover, we demonstrate that such interface states can also be generated in hybrid structures that combine two superlattices with bandgaps of different orders centered around the same frequency. These structures open up new avenues for exploring topological confinement in high-order bandgaps, providing a unique platform for unveiling and better understanding complex topological systems.
\end{abstract}

\section{\label{sec:level1}Introduction}

A periodic lattice containing two elements per unit cell can be described by the one-dimensional Su-Schrieffer-Heeger (SSH) model in the tight-binding approximation. This description has been a significant breakthrough for developing materials with topological properties \cite{su_solitons_1979, asboth_short_2016}. Topological states have since been demonstrated for a wide variety of excitations, including photons \cite{ozawa_topological_2019, hauff_chiral_2022, hafezi_imaging_2013, xiao_surface_2014, lu_topological_2014, khanikaev_two-dimensional_2017}, phonons \cite{esmann_topological_2018, arregui_coherent_2019,PhysRevB.97.020102, susstrunk_observation_2015, prodan_topological_2009}, vibrations \cite{doster_observing_2022, ma_topological_2019,xiao_geometric_2015,zhao_topological_2018,zheng_observation_2019,zhang_topological_2018,he_acoustic_2016,huber_topological_2016}, polaritons \cite{st-jean_measuring_2021, pernet_gap_2022, klembt_exciton-polariton_2018}, plasmons \cite{song_plasmonic_2021}, and magnons \cite{malz_topological_2019}. In the context of periodic lattices, multilayered structures, such as distributed Bragg reflectors (DBRs), have high-reflectivity regions associated with bandgaps. The states at the edge of these bandgaps present different spatial symmetries \cite{jusserand_raman_1989}. When concatenating two DBRs with inverted spatial mode symmetries around a gap, a topological interface state emerges \cite{xiao_surface_2014, esmann_topological_2018_1,esmann_topological_2018_2}.

Nanophononics~\cite{priya_perspectives_2023, ortiz_fiber-integrated_2020, lanzillotti-kimura_nanowave_2006}, i.e., the engineering of acoustic nanowaves, appears as a versatile simulation platform~\cite{esmann_topological_2018,ortiz_phonon_2019}. Unlike in optics or electronics \cite{parappurath_direct_2020,lateral_Bragg_Papatry, Wang_2018}, the linear dispersion relation of acoustic phonons \cite{delsing_2019_2019} allows for the study of topological interface states in a broad frequency range. In particular, nanoacoustic topological states have been evidenced in superlattices working at acoustic frequencies in the tens to hundreds of GHz range, and demonstrated exceptional agreement between theory and experiments.~\cite{ortiz_topological_2021, esmann_brillouin_2019} In these reported cases, the unit cell was formed by two materials whose thickness ratio was optimized to reverse the mode symmetries around a specific bandgap while keeping the acoustic thickness constant. These studies demonstrated the robustness of the topologically protected states against thickness perturbations that do not affect the Zak phase ($\theta^{Zak}$) \cite{ortiz_topological_2021, esmann_brillouin_2019}, which is a key parameter to characterize topological phases in the SSH model \cite{zak_berrys_1989}. Formally, the Zak phase is defined as the integral of the displacement across the Brillouin zone, and it can be associated with the sign of the reflection phase of a finite-size DBR \cite{xiao_surface_2014, xiao_geometric_2015}. Each individual band of the acoustic dispersion relation has an associated $\theta^{Zak}$, which can take only two values, 0 or $\pi$. In this work, we theoretically investigate topological interface states by concatenating DBRs with inverted spatial mode symmetries at high-order bandgaps. Based on the different Zak phase configurations for the different bandgaps at specific unit cell thickness ratios, we engineer interface modes at higher bandgap orders. Furthermore, we benefit from the linear dispersion relation to generate hybrid topological resonators. We establish the interface states of these resonators by concatenating two superlattices with different bandgap orders that are overlapping in frequency. We thus show that the presence of the topological states does not depend only on the spatial mode symmetry but also on the relative bandgap order of the concatenated superlattices. This platform allows to experimentally map textbook cases, and test innovative and counterintuitive physical situations. Our findings unlock a new degree of freedom to design topological nanoacoustic resonators.

The paper is organized as follows. Section \ref{sct:multimode} describes the principle of band inversion in the context of nanoacoustics. In Section \ref{sct:interface high order}, we present the method for generating interface states at high-order bandgaps and discuss their robustness compared to Fabry-Perot resonators. Sections \ref{sct:multimode engineering} and \ref{sct:hybrid resonator} introduce our new designs of topological acoustic resonators.

\section{Principle of Band Inversion} \label{sct:multimode}

In topological superlattices, band inversion refers to the situation where two modes with opposite symmetries at the edges of a bandgap exchange their ordering in energy. This inversion can be achieved by adjusting the relative thickness of the two layers in the unit cell of the superlattice. By concatenating two superlattices with different layer thicknesses, i.e., superlattices presenting inverted bands, topologically protected interface states can be created.

\begin{figure}[!ht]
\centering
\includegraphics[scale = 0.82]{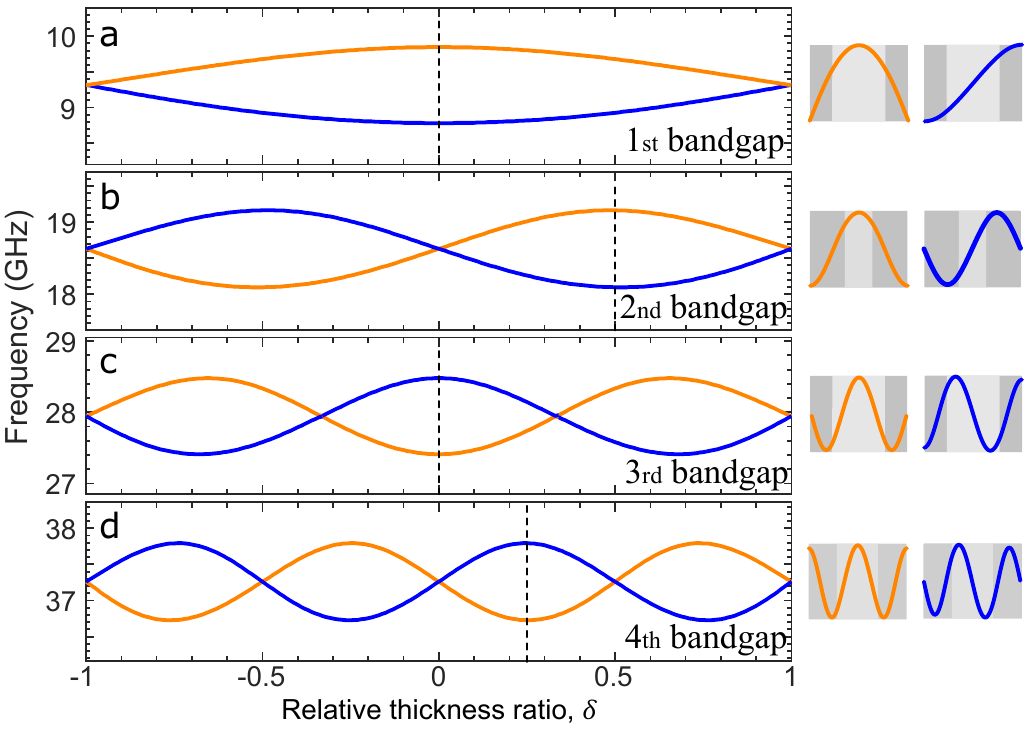}
\caption{Left: Frequency of the band-edges bounding the bandgap as a function of the parameter $\delta$. (a)-(d) Band inversion of the acoustic bandgap around 9.3 GHz (a), 18.6 GHz (b), 28 GHz (c), and 37.3 GHz (d). The mode symmetries are indicated by orange (symmetric) and blue (anti-symmetric) lines. The thickness ratio of GaAs/AlAs is indicated by the vertical dashed line in each case. Right: The unit cell is displayed with light and dark gray colors, representing AlAs and GaAs, respectively. The band-edge modes are shown in the unit cells for each bandgap.}
\label{fig:BndIversion-multimode}
\end{figure}

In the case of multilayered GaAs/AlAs structures composed of two concatenated distributed Bragg reflectors (DBRs), we introduce a parameter $\delta \in [-1, : 1]$ to represent the relative thickness of AlAs and GaAs.~\cite{esmann_topological_2018_1} In DBRs with a centro-symmetric unit cell centered around the AlAs layer, GaAs is distributed equally on both sides of AlAs as follows: $\frac{\lambda_{GaAs}}{8}(1+\delta)$, $\frac{\lambda_{AlAs}}{4}(1-\delta)$, $\frac{\lambda_{GaAs}}{8}(1+\delta)$. Figure~\ref{fig:BndIversion-multimode} illustrates the evolution of the first four bandgaps as a function of $\delta$, while maintaining the unit cell acoustic thickness constant. The orange lines indicate modes symmetric with respect to the center of the unit cell, while the blue lines indicate anti-symmetric modes. It can be observed that the modes exhibit a sinusoidal dependence on the parameter $\delta$. Starting from the second bandgap, there is a consecutive opening and closing of the bandgap as $\delta$ increases, accompanied by an inversion of symmetry. The number of nodes in the modes is directly related to the order of the bandgap. The first bandgap (Fig.\ref{fig:BndIversion-multimode}(a)) opens and closes only once, with a maximum amplitude at $\delta = 0$. The second bandgap (Fig.\ref{fig:BndIversion-multimode}(b)) opens twice and closes at $\delta = 0$, with a symmetry inversion of the edge modes around this point. The third bandgap exhibits two symmetry inversions (Fig.\ref{fig:BndIversion-multimode}(c)), while the fourth gap undergoes three symmetry changes across the full range of $\delta$ (Fig.\ref{fig:BndIversion-multimode}(d)). Generally, the $n^{th}$ bandgap experiences $(n-1)$ symmetry inversions.

The topological properties of a multilayered acoustic device can be characterized using the Zak phase, which is the 1D equivalent of the Berry phase \cite{berry_original}. For a periodic phononic 1D system with a periodicity $a$, the Zak phase of the $n^{th}$ band is calculated by integrating across the Brillouin zone:

\begin{equation}
\theta_n^{Zak}=\int_{-\pi/a}^{\pi/a} \left[ i \int_{\text{unit cell}} \frac{1}{2\rho(z)v^2(z)} dzu^*{n,k}(z)\partial_ku{n,k}(z) \right]dk,
\end{equation}
where $u_{n,k}(z)$ is the acoustic displacement of the $n^{th}$ band and wave-vector $k$ at position $z$, and $\rho(z)$ and $v(z)$ correspond to the mass density and speed of sound in the materials.~\cite{xiao_geometric_2015, zak_berrys_1989, esmann_topological_2018_1}

In a periodic system with inversion symmetry, where the unit cell is centro-symmetric around AlAs, the Zak phase can only take on two discrete values: 0 or $\pi$ \cite{xiao_surface_2014, xiao_geometric_2015}. It is associated with the symmetries of the Bloch modes at the band edge \cite{xiao_surface_2014}. When the modes at both ends of the same $n^{th}$ band (i.e., at the edge and center of the Brillouin zone) have the same symmetries, the Zak phase $\theta_n^{Zak}$ for that band is 0. Conversely, if the band has edge modes with opposite symmetries, $\theta_n^{Zak} = \pi$ \cite{xiao_surface_2014}. This parameter is crucial for predicting interface states and characterizing one-dimensional topological systems.

\section{Interface states at high-order bandgaps} \label{sct:interface high order}
Fundamentally, an interface state is formed whenever two DBRs with opposite reflection phase signs are concatenated~\cite{esmann_topological_2018_1, xiao_surface_2014}. The reflection phase can be either positive or negative depending on the structure of the DBR. For the $n^{th}$ bandgap, one can determine the sign of the reflection phase by evaluating the relation~\cite{xiao_surface_2014}:
\begin{align}
    sgn(\phi) = (-1)^n(-1)^l \times \exp{\left( i\sum_{m=0}^{n-1}\theta_m^{Zak}\right)},
\end{align}
where $l$ is the number of closed bandgaps below the $n^{th}$ bandgap.  Therefore, for the $n^{th}$ bandgap, there is an interface state at the condition that $\sum_{m=0}^{n-1}\theta_m^{Zak} = 0 + 2p\pi, p \in \mathbb{N}$ for one DBR and $\sum_{m=0}^{n-1}\theta_m^{Zak} = \pi + 2p\pi, p \in \mathbb{N}$ for the other. The creation of an interface state at the $n^{th}$ bandgap does not necessarily imply the generation of an interface state in other bands, as the sum of $\theta^{Zak}$ for each band might differ.

\begin{figure}[h]
    \centering
    \includegraphics[scale = 0.7]{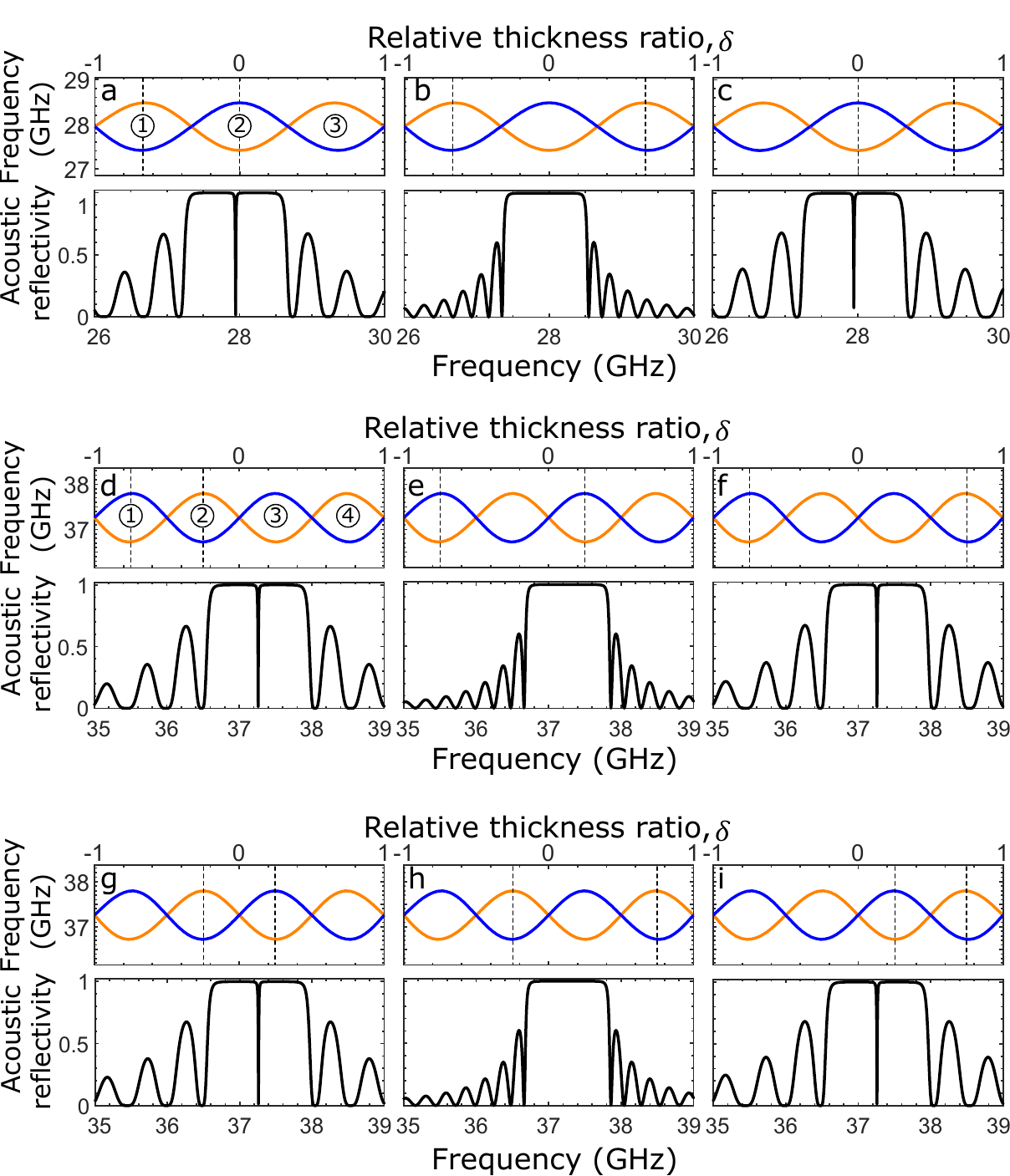}
    \caption{Top panels: Band inversion of the (a)-(c) third acoustic bandgap around 28 GHz, and (d)-(i) fourth bandgap around 37.3 GHz. Bottom panels: calculated acoustic reflectivity spectra for the concatenated DBRs with the corresponding GaAs/AlAs thickness ratio $\delta$  marked as dashed lines on the top panels.}
    \label{fig:band-inversion-3rd}
\end{figure}

Figures~\ref{fig:band-inversion-3rd} (a)-(c) display reflectivity spectra for the third bandgap obtained by combining two DBRs with different values of $\delta$, indicated by the dashed lines on the corresponding top panels. In each case, $\delta$ is chosen such that the amplitude of the bandgap of the corresponding superlattice is maximized, which corresponds to the values $\delta=0$ and $\pm0.66$. In panels (a) and (c), the two DBRs have inverted symmetry around the bandgap. In both cases, the acoustic reflectivity contains a dip centered in the high reflectivity region featuring the interface state. In contrast, in panel (b), the two DBRs have the same symmetry around the bandgap. Thus, this structure acts as a standard DBR, with a high reflectivity region. Generally, for the third bandgap, by concatenating one DBR with $\delta\in [-0.33, 0.33]$ and another with $\delta < 0.33$ (Fig.\ref{fig:band-inversion-3rd}(a)) or $\delta > 0.33$ (Fig.\ref{fig:band-inversion-3rd}(c)), the band inversion is preserved, and an interface state is generated.

Regarding the fourth bandgap, Figs.~\ref{fig:band-inversion-3rd} (d)-(i) show six possible combinations between the two DBRs. In all shown cases, the values of $\delta$ are chosen such that the amplitude of the bandgap is maximized, resulting in high reflectivity regions centered at $\sim$37.3 GHz. Depending on the mode symmetries around the fourth gap, an interface state is either present or absent. In the cases shown in Figs.~\ref{fig:band-inversion-3rd}(e) and (h), the modes of the two concatenated DBRs have the same symmetry (both modes at the bottom band are symmetric in panel (e), while they are anti-symmetric in panel (h)). Thus, the acoustic reflectivity spectra present high reflectivity regions like a standard DBR. On the contrary, in Figs.~\ref{fig:band-inversion-3rd} (d), (f), (g), and (i), the two DBRs that are concatenated have inverted mode symmetries. As a result, an interface state between the two concatenated DBRs is generated. Generally, the approach presented for the third and fourth bandgaps can be extended to higher bandgap orders. More specifically, an interface state is generated when concatenating any two DBRs corresponding to superlattices with an even and odd bandgap opening.

These interface states can be accessed experimentally through Brillouin scattering measurements.~\cite{arregui_coherent_2019,ortiz_topological_2021} Nevertheless, the scattering cross-section ($\sigma$), which represents the magnitude of the scattered signal, relies on the relative thickness of GaAs and AlAs within the two unit cells constituting each superlattice. Consequently, not all theoretically predicted interface states can be accessed experimentally. To identify the superlattice combinations that yield experimentally accessible interface states, we employed the transfer matrix method and a photoelastic model to simulate the Brillouin cross-section of the interface states.~\cite{lanzillotti-kimura_phonon_2007,lanzillotti-kimura_theory_2011} The Brillouin cross-section is defined by the overlap integral between the incident laser electric field $E(z)$, the strain, which is given by the derivative of the displacement $\frac{\partial u(\omega,z)}{\partial z}$, and the photoelastic constant $p(z)$ over the whole structure in the form:
\begin{equation}
    \sigma(\omega) = \int |E(z)|^2p(z)\frac{\partial u(\omega,z)}{\partial z}dz,
    \label{eq:cross-section}
\end{equation}
where $p(z)$ is material dependent, being $p = 1$ ($p = 0$) in Ga-rich (Al-rich) layers~\cite{ortiz_topological_2021}. We considered here the whole acoustic structure as an optical $\lambda$-cavity embedded in vacuum, where $\lambda_\textrm{opt} \sim1600$ nm. The cross-section depends on the overlap between the electric and acoustic fields $|E(z)|^2(\partial u(\omega,z)/\partial z)$ in the regions where the photoelastic constant distribution $p(z)$ is non-zero \cite{ortiz_topological_2021}.

\begin{figure}
    \centering
    \includegraphics[scale = 0.7]{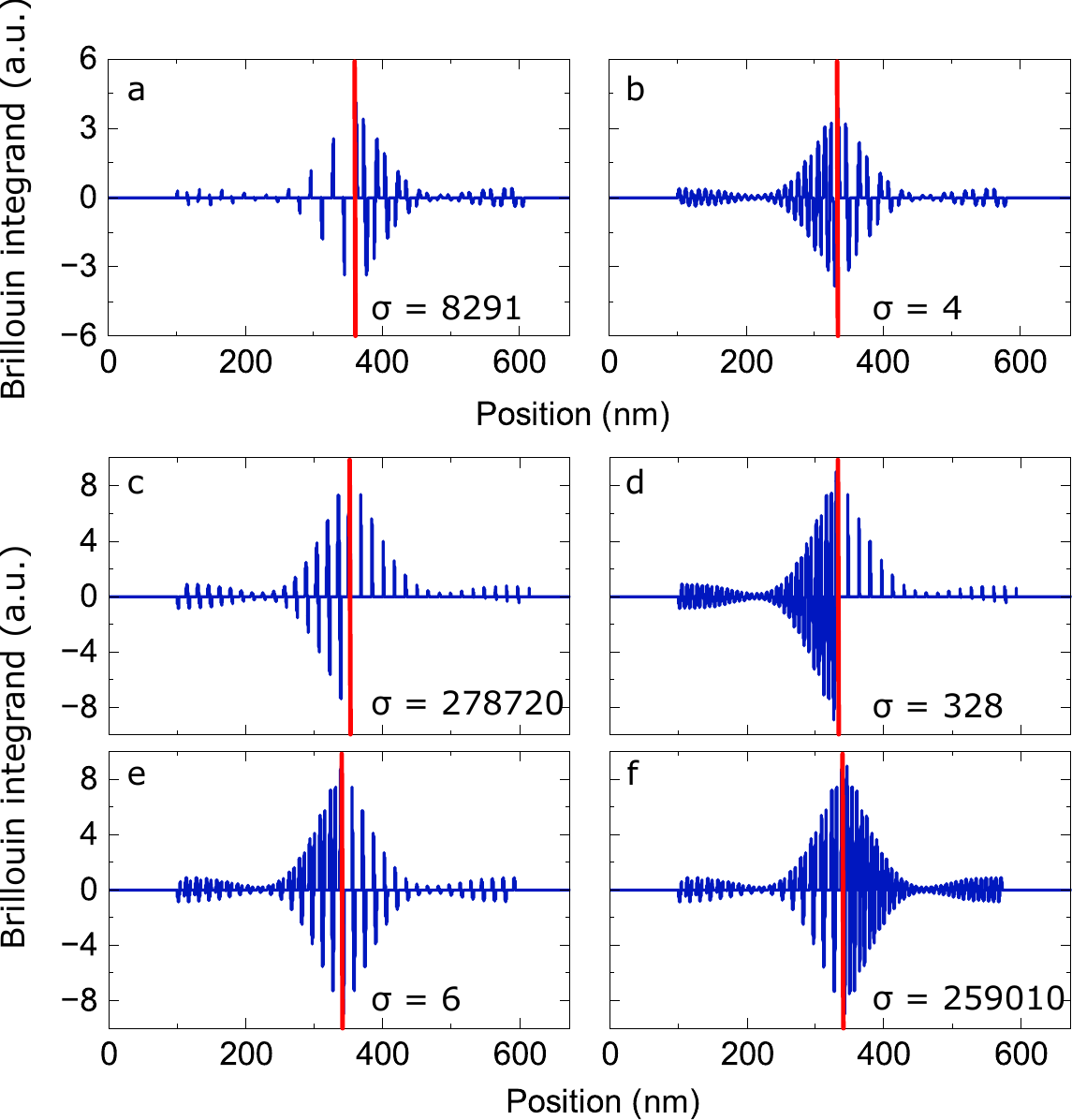}
    \caption{Integrand of the Brillouin cross-section for interface states in the  third bandgap (a),(b), associated to the structures shown in Figs.~\ref{fig:band-inversion-3rd}(a) and (c), respectively; and the fourth bandgap (c)-(f), associated to the structures of Figs.~\ref{fig:band-inversion-3rd}(d), (f), (g) and (i), respectively. The vertical red line indicates the interface between the two DBRs.}
    \label{fig:integrand}
\end{figure}

The integrand of Eq.~\ref{eq:cross-section} is displayed in Fig.~\ref{fig:integrand} for structures supporting interface states in the 3rd and 4th bandgaps. We can analyze the integrand by splitting it into quadrants: left/right superlattice and positive/negative amplitude contributions. The integrand features signals composed of either double peaks (thick lines) or single peaks (thin lines). Figs. \ref{fig:integrand} (a) and (b), display the integrand of the modes in the third bandgap, associated with the topological structures presented in Figs.~\ref{fig:band-inversion-3rd}(a) and (c), respectively. The peaks with the maximum amplitude at the interface between the two DBRs are the main contributors to the overall cross-section, resulting in a high Brillouin cross-section in panel (a). Conversely, the positive and negative contributions of the integrand displayed in panel (b) cancel each other, leading to a low Brillouin cross-section. The calculated Brillouin cross-sections for the mode in the third bandgap on both cases are, respectively, $\sigma = 8291$ and $\sigma = 4$.

Figs.~\ref{fig:integrand}(c)-(f) display the integrand of the modes in the fourth bandgap, associated to the cases of Figs.~\ref{fig:band-inversion-3rd}(d), (f), (g), and (i), respectively. The integrand in Figs.~\ref{fig:integrand}\textbf(c) and (f) exhibit positive contribution across the entire structure, while negative amplitude contribution is asymmetric between left and right quadrants. This results in an overall positive signal with high Brillouin cross-sections of $\sigma = 278720$ and $\sigma = 259010$, respectively. In contrast, the Brillouin cross-section in Figs.~\ref{fig:integrand}(d) and (e) is smaller; $\sigma = 328$ and $\sigma = 6$, respectively. A zoomed-in version of the graphs depicting the integrands in more detail is shown in the Supplementary Information. We note that although the apparent overall contribution is positive, the positive right quadrant contribution in panel (d) is not substantial, resulting in a fairly small cross-section. Similarly, in panel (e), even though the positive left quadrant displays double peaks, the negative peaks are thicker than the positive contributions, resulting in a small Brillouin cross-section as well. 

Topological interface states have been demonstrated to exhibit exceptional robustness against disorder, which is useful for transport and error-free data communication.~\cite{mathew_synthetic_2020,arora_direct_2021, Data_Comm_L3Matrix}. The band inversion principle exploited here to build topological resonators preserves the center of the bandgap when varying $\delta$. We have numerically demonstrated that the robustness of the interface mode applies to all bandgap orders when introducing fluctuations in the layer thickness ratio. In practice, such fluctuations might emerge, for example, by material intermixing or composition fluctuations~\cite{RefInd,Qdash} during the epitaxial growth of the layers that form the superlattice. The fluctuations considered here, concern changes in the GaAs/AlAs ratio while maintaining a constant unit-cell acoustic thickness. In the model, we implement this by using a flat distribution of random numbers with an amplitude $\Delta\delta/\delta$ ranging from zero (unperturbed system) to 0.5. We note that a noise amplitude set to 0.999 effectively closes the targeted bandgap for each designed superlattice.

\begin{figure}[ht]
    \centering
    \includegraphics[scale = 0.7]{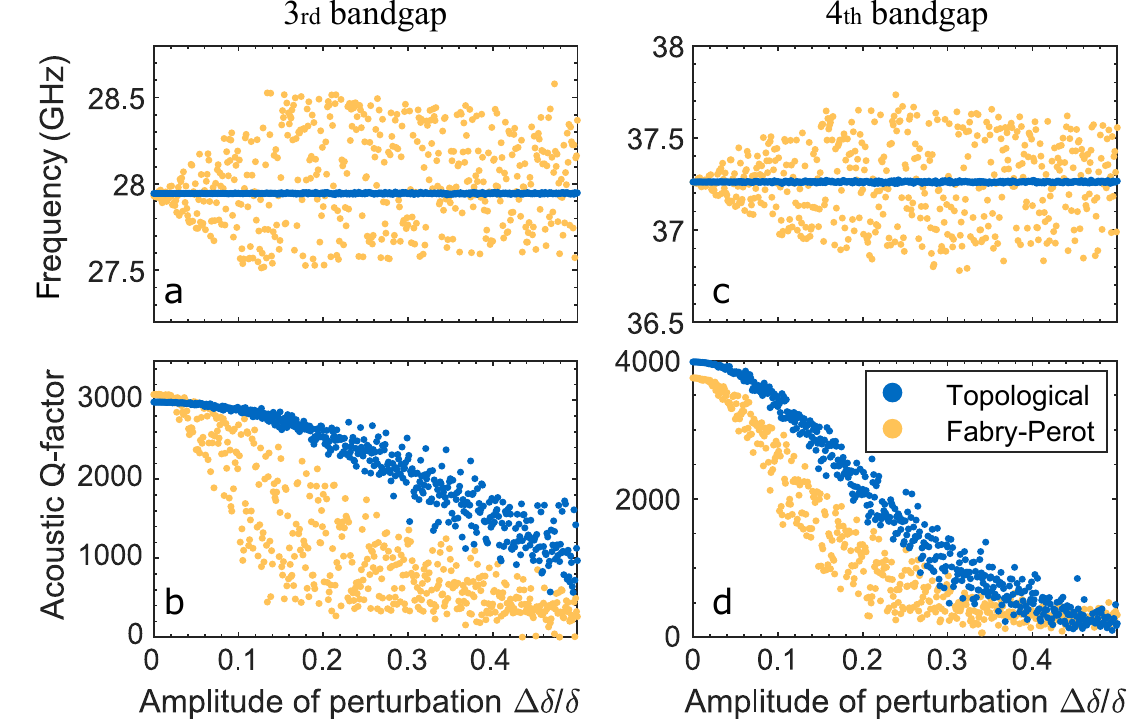}
    \caption{(a) Resonant frequency in the third bandgap under random perturbations, with a uniform distribution of width $\Delta\delta/\delta$. The acoustic resonance frequency stays trapped at the bandgap center for the topological mode (blue) but undergoes variations in the Fabry–Perot (orange). (b) Acoustic quality factor under random perturbations. For both types of resonators, the acoustic quality factor drops by a factor of ten. Similarly, (c) and (d) display the same comparisons for the fourth bandgap.}
    \label{fig:robustness_multi}
\end{figure}

In Figs.~\ref{fig:robustness_multi}(a) and (b), we compare the topological interface state generated in the third bandgap (corresponding to the structure shown in Fig.~\ref{fig:band-inversion-3rd}(c)) to the confined mode in a Fabry-Perot (FP) cavity. The FP cavity considered here is formed by non-centrosymmetric unit cells of GaAs/AlAs with $\delta = 0.66$, chosen such that it preserves the maximum bandgap opening. The two DBRs surround a spacer of thickness $\lambda$, and are embedded in a GaAs background. 
Figure~\ref{fig:robustness_multi}(a) displays the acoustic frequencies of the topological (blue) and the Fabry-Perot (orange) resonators as a function of $\delta$-fluctuations. In the FP resonator, the fluctuations are introduced in the thickness ratio of GaAs and AlAs constituting the DBR unit cell, while the spacer has no perturbation. In this case, the topological resonator maintains its resonance at the bandgap center. In contrast, the resonance frequency of the FP resonator undergoes large fluctuations, spanning up to 0.5 GHz away from the center frequency. Figure~\ref{fig:robustness_multi}(b) shows the influence of thickness fluctuations on the acoustic quality factors of the two structures. As seen, both quality factors decrease by a factor of 10 at high perturbation amplitudes. This effect can be explained by the effective reduction in the width of the bandgap, which results in an increase of the evanescent decay length of the confined mode, and so an enhancement of the leakage through the DBRs towards the background~\cite{ortiz_topological_2021}. Despite the decrease in quality factors for both structures with increasing perturbation amplitudes, the topological structure shows more consistent Q-factors compared to the FP resonator. This is related to the fact that the fluctuations in the topological mode frequency are less prominent, and so its associated Q-factor decreases at a slower pace.

Likewise, in Figs.~\ref{fig:robustness_multi}(c) and (d), we compare the robustness of the topological interface state generated in the fourth bandgap to an FP cavity ($\delta = 0.75$), with a spacer of thickness $\lambda$/2. Figure~\ref{fig:robustness_multi}(c) shows the resonant frequencies of both resonators as a function of the fluctuations. The observed behavior is similar to the case of the third bandgap: the frequency is clamped at the bandgap center for the topological structure and fluctuates for the Fabry-Perot resonator. These results show that the robustness characteristic of topological devices, protecting the acoustic resonance against disorder, is also preserved at high bandgap orders. However, as we see by comparing Figs.~\ref{fig:robustness_multi}(b) and (d), the acoustic quality factor of the fourth bandgap (Fig.~\ref{fig:robustness_multi}(d)) in both structures (FP and topological) is more sensitive to fluctuations in comparison to the third bandgap (Fig.~\ref{fig:robustness_multi}(b)). This can be understood intuitively by comparing the opening and closing of the bandgaps as a function of unit cell composition, as shown, for example, in Fig.~\ref{fig:BndIversion-multimode}. When transitioning to higher-order bandgaps, the opening/closing of each bandgap necessitates gradually smaller adjustments in material thicknesses. Consequently, interface states at higher-order bandgaps become progressively more susceptible to inaccuracies in material thicknesses.

\section{Multimode engineering} \label{sct:multimode engineering}

In the previous section, we showed that we could engineer the interface states at the $n^{th}$ bandgap in topological acoustic resonators by carefully tuning the unit-cell material thickness ratios in both juxtaposed DBRs. In this section, we will show that we can also generate interface states at multiple bandgap orders simultaneously. By varying the unit-cell relative thickness ratio, $\delta$, the bandgap amplitudes at all the orders are, in fact, simultaneously altered. However, the closing and reopening of the bandgaps are not coincident for every bandgap order. As a result, by changing $\delta$, we can reach different combinations of bandgap symmetries, and so we can engineer the formation of interface states at different bands. 

\begin{figure*}
    \centering
    \includegraphics[scale = 0.7]{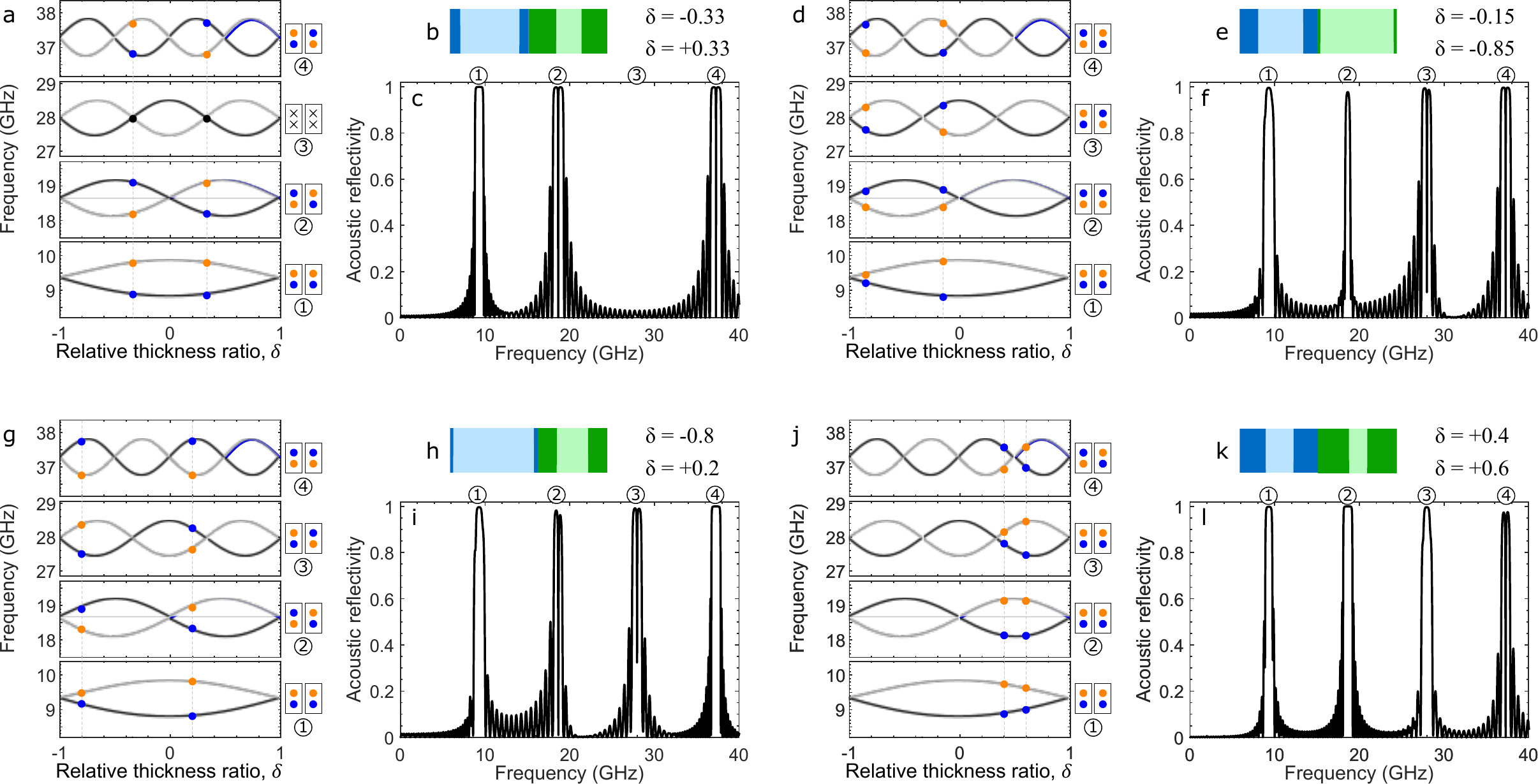}
    \caption{Engineering of topological interface states. (a) Band inversion of the acoustic bandgaps. The dots show the bandgaps opening and symmetries at the given $\delta$. Inset: Symmetry of the two DBRs concatenated. There is inversion only for the fourth bandgap. (b) Schematic of the unit cell configurations at the interface between the two DBRs, with the dark blue and green colors representing GaAs and the corresponding light colors representing AlAs. (c) Simulated acoustic reflectivity spectra for two topological resonators formed by two DBRs concatenated embedded in GaAs with $\delta_{top} = -0.33$ and $\delta_{bottom} = +0.33$. Likewise in
   (d)-(f) $\delta_{top} = -0.15$ and $\delta_{bottom} = -0.85$; (g)-(i) $\delta_{top} = -0.8$ and $\delta_{bottom} = +0.2$; and (j)-(l)  $\delta_{top} = +0.4$ and $\delta_{bottom} = +0.6$. }
    \label{fig:reflectivity-multimode-one-for-all}
\end{figure*}

Figure~\ref{fig:reflectivity-multimode-one-for-all} presents different conditions to generate interface states at different gaps. In the first case (Figs.~\ref{fig:reflectivity-multimode-one-for-all}(a)-(c)), we optimize the GaAs/AlAs thickness ratio to generate interface states at the second and fourth bandgaps, with $\delta = \pm0.33$ in each DBR. Fig.~\ref{fig:reflectivity-multimode-one-for-all}(a) shows the dependence of the bandgaps on $\delta$, with the dashed vertical lines indicating the $\delta$ values of each DBR ($\delta = \pm0.33$), and the orange (blue) dots corresponding to the symmetric (anti-symmetric) modes. Fig.~\ref{fig:reflectivity-multimode-one-for-all}(b) shows the unit cell configurations at the interface for the chosen $\delta$, in which the dark blue and green colors represent GaAs whereas light blue and green represent AlAs. The inset at the right side of Fig.~\ref{fig:reflectivity-multimode-one-for-all}(a) shows the mode symmetry of the two DBRs in each band. As shown,  at the particular value of $\delta$ chosen here, there is an inversion of symmetry at the second and the fourth bandgap, whereas the third bandgap is closed (see black dots). In the calculated acoustic reflectivity spectrum shown in Fig.~\ref{fig:reflectivity-multimode-one-for-all}(c), three bandgaps can be seen. In these bandgaps, the band-inversion interface states are present for the 2nd and 4th gaps, as indicated by the dips centered in the high reflectivity regions at $\sim$18 GHz and $\sim$37 GHz, respectively. As we already saw in Fig.~\ref{fig:reflectivity-multimode-one-for-all}(a), the third gap is closed for this combination of  $\delta$s in the DBRs, and so that gap is absent in the reflectivity spectrum.

Figures \ref{fig:reflectivity-multimode-one-for-all}(d)-(f) present the conditions that generate interface states in the third and fourth bandgap simultaneously. The structure is designed with $\delta_{top/bottom} = -0.15 /-0.85$ for the top and bottom DBRs. In comparison to the previous case, the four bandgaps are open for both superlattices (see panel (d)), resulting in four high reflectivity regions in the reflectivity spectrum, as shown in panel (f). For this configuration of unit cells, interface states are present in the third and fourth bandgaps. They are induced by the inversion of symmetry of the modes around the respective bandgaps, as shown in panel (d). On the contrary, there is no interface mode in the first and second bandgaps, as they have the same band-edge symmetries for both superlattices.

We note that the mode is not centered in the third bandgap, as shown in Fig. \ref{fig:reflectivity-multimode-one-for-all}(f). To generate an interface that is centered in the bandgap, there are two necessary conditions. First, the two DBRs should have a bandgap with equal central frequency and, second, they should have the same bandwidth. The first condition is required for topological robustness~\cite{ortiz_topological_2021}. The second condition results in similar evanescent decay lengths into both DBRs. In case the values of $\delta$ for each superlattice are not equidistant from a band inversion point, the interface state generated is not centered in the bandgap, as we can see in Figs.~\ref{fig:reflectivity-multimode-one-for-all}(f) and (i) at the third and second bandgaps, respectively. 
Figures~\ref{fig:reflectivity-multimode-one-for-all}(g)-(i) present the conditions to generate interface states in the second and third bandgaps, associated to $\delta_{top/bottom} = -0.8 /+0.2$. In Fig.~\ref{fig:reflectivity-multimode-one-for-all}(g) the two DBRs have inverted symmetry at the targeted bandgaps, whereas the mode symmetries at the fourth bandgap on both DBRs are the same, even though they fall into different bandgap openings. This results in two interface states at bandgap orders 2 and 3, as seen in Fig.~\ref{fig:reflectivity-multimode-one-for-all}(i). Figures~\ref{fig:reflectivity-multimode-one-for-all}(j)-(l) present the conditions to generate interface states only in the fourth bandgap, with $\delta_{top/bottom} = +0.4 /+0.6$. Despite the slight disparity between the unit cells of the two superlattices in this arrangement, the interface state is generated (Fig.~\ref{fig:reflectivity-multimode-one-for-all}(k)). Only the modes at the fourth bandgap have inverted symmetries, whereas all the other gaps exhibit the same mode symmetry. As a result, the acoustic reflectivity spectrum, displayed in  Fig.~\ref{fig:reflectivity-multimode-one-for-all}(l), presents three high reflectivity regions and one interface mode in the fourth bandgap.

The first bandgap does not undergo any symmetry inversion of the topological phase over the entire $\delta$ range, which makes it impossible to create an interface state at this band. It is important to point out that this does not mean that interface states cannot be conceived in the first bandgap. For instance, one can tune the impedance of the materials to switch the symmetry of the modes.~\cite{xiao_surface_2014, esmann_topological_2018_2}

\section{Hybrid topological resonators} \label{sct:hybrid resonator}

The formation of interface states is not limited to the same bandgap order of the constituting DBRs. In fact, the general rule for engineering topological states is associated with the sign of the reflection phase as well as the overlap of the high reflectivity regions from both reflectors. So far, we have fulfilled these conditions and created interface modes at higher-order bandgaps by concatenating two DBRs designed at the same acoustic frequency. In this section, we extend this concept to generate topological states between two DBRs designed at different fundamental frequencies, resulting in bandgaps of different orders sharing the same frequency range.

\begin{figure*}
    \centering
    \includegraphics[scale = 0.92]{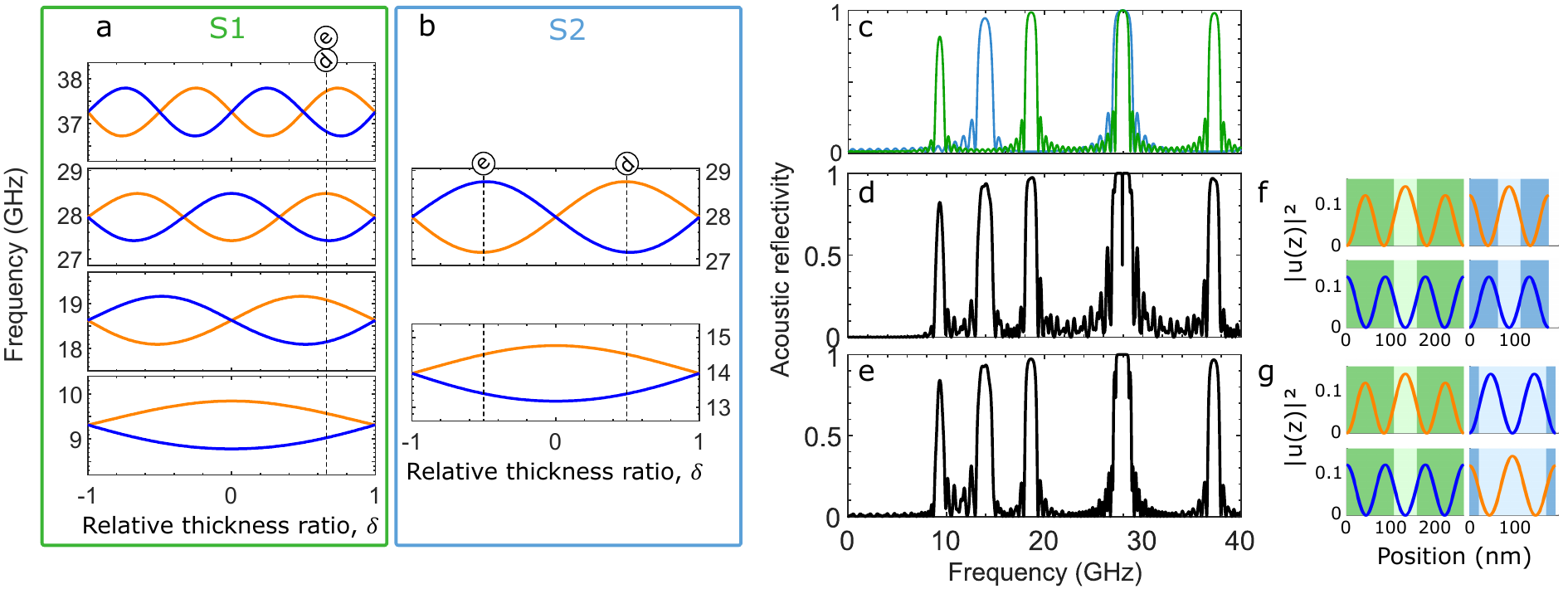}
    \caption{Hybrid topological acoustic resonator. (a),(b) Band inversion of the acoustic bandgaps associated to the two concatenated superlattices, named (a) S1 (green frame), and (b) S2 (blue frame). The third bandgap of S1 and the second bandgap of S2 share the same central frequency, around 28 GHz. (c) Simulated acoustic reflectivity of the two DBRs. The green (blue) line corresponds to the band inversion diagram displayed in panel (a) (panel (b)). (d),(e) Simulated acoustic reflectivity spectra for different concatenated DBRs embedded in GaAs. The relative thickness ratio $\delta$ of the corresponding DBRs is marked by dashed lines on panels (a) and (b). (f),(g) Acoustic displacement $|u(z)|^2$ of the modes at the edge of the bandgaps plotted on top of the unit cell for the two superlattices. The green (blue) unit cell corresponds to the band inversion diagram displayed in panel (a) (panel (b)), and the dark (light) colors represent GaAs and AlAs, respectively.}
    \label{fig:hybrid-band-inversion}
\end{figure*}

Figures~\ref{fig:hybrid-band-inversion}(a) and (b) display the band inversion diagrams of two superlattices, S1 and S2, intended to have the first bandgap at different frequencies. The first superlattice (S1) is designed to have a fundamental bandgap centered $\sim$9.3 GHz (Fig.~\ref{fig:hybrid-band-inversion}(a)), while the first bandgap of S2 is centered at $\sim$14 GHz (Fig.~\ref{fig:hybrid-band-inversion}(b)). These designs result in the third bandgap of S1 centered at the same frequency as the second bandgap of S2, which is represented by the alignment of both bands in panels (a) and (b). Fig.~\ref{fig:hybrid-band-inversion}(c) shows the acoustic reflectivity spectra of the two DBRs associated with the band structures in panels (a) and (b). The reflectivities are calculated for values of $\delta$ in which all the bandgaps are open. We see that there is a complete overlap of the high reflectivity regions around 28 GHz. By concatenating two such superlattices with overlapping bandgaps of different order, we can generate an interface state at the frequency in which they overlap.

Figs.~\ref{fig:hybrid-band-inversion}(d) and (e) depict the calculated acoustic reflectivity spectra for two different combinations of such DBRs. The chosen values of $\delta$ are marked on the band inversion figures (Figs.~\ref{fig:hybrid-band-inversion}(a) and (b)) with the vertical dashed lines. The matching labels between panels (a) and (b) indicate the corresponding acoustic reflectivity spectrum. The first superlattice on both structures is the same, where $\delta=0.66$ corresponds to the thickness ratio for which the third bandgap opening is maximized. On the other hand, the thickness ratio of the second superlattice is chosen to maximize the second bandgap, where $\delta=0.5$ ($\delta=-0.5$) correspond to the same (inverted) band edge mode symmetries compared to the third order bandgap of S1. In both cases (panels (d) and (e)), there are five high-reflectivity regions. The regions below 25 GHz and above 35 GHz correspond to the individual bandgaps of S1 and S2 that have no overlap. At 28 GHz, different bandgaps of the two DBRs do overlap, but an interface state is either present (panel (d)) or absent (panel (e)) in the two shown examples. Contrary to what one might have expected based on the results of the previous sections, the combination of superlattices supporting an interface state (panel (d)) have the same band edge mode symmetries. Likewise, when the symmetry of the modes between S1 and S2 is inverted, no interface state is generated. Therefore, the rule of creating an interface state by band inversion cannot be blindly applied for bandgaps of different orders. 

To understand this, we investigate the relation between the generation of an interface state and the acoustic displacement field in the unit cell of each superlattice. In Fig.~\ref{fig:hybrid-band-inversion}(f) and (g), we can see a schematic of the unit cells of superlattices S1 and S2, associated with the acoustic reflectivity spectra shown in Figs.~\ref{fig:hybrid-band-inversion}(d) and (e). As before, the dark- and light-colored regions correspond to GaAs and AlAs, respectively. On top of these superlattices, we show the corresponding relevant acoustic displacements. More specifically, in the top (bottom) panels of this schematic, we show the acoustic displacement of the modes at the higher (lower) frequency band edges of the bandgap centered at 28 GHz in each superlattice. As we can see in Fig.~\ref{fig:hybrid-band-inversion}(f), there is a discontinuity of the displacement field at the interface between these superlattices. This discontinuity leads to the generation of an interface state, even though the band edge modes have the same symmetry. On the other hand, in Fig. \ref{fig:hybrid-band-inversion}(g), the displacement field at the interface between S1 and S2 is continuous, and so prevents the formation of an interface state. In general, when the order of one bandgap is even, and the other one is odd, an interface state is generated if both have the same symmetries. On the contrary, if two odd or even bandgaps are concatenated, they must have inverted symmetries to generate an interface state. The difference in phases of the reflection coefficient does not only depend on the sum of the Zak phases anymore but also on the order of each bandgap. The case of two DBRs with different lattice parameters was briefly considered in Ref.~\cite{xiao_surface_2014}.

\section{Conclusions}
We theoretically presented a method to generate acoustic interface states in topological nanoacoustic resonators based on the band inversion principle. We simulated a series of topological optophononic resonators with different combinations of concatenated superlattices. By changing the thickness ratio of GaAs and AlAs in the unit cell, we were able to control the symmetries of the modes around each bandgap. In general, an interface state can be generated when two superlattices with inverted symmetries are concatenated. Here, we extended this principle to create interface states in high-order bandgaps. The modes that we presented can be accessed experimentally in a Brillouin or pump-probe experiment.  We numerically discussed the Brillouin efficiency of different combinations of superlattices forming topological states at the third and fourth bandgap orders. The accessibility of these states in Brillouin scattering experiments is directly associated with the unit cell thickness ratio between GaAs and AlAs of both concatenated superlattices. In addition, we studied the robustness of our structures against disorder and compared them with Fabry-Perot resonators. The use of GaAs and AlAs enables the study of electronic and optical resonances effects,~\cite{lanzillotti-kimura_enhanced_2011-1, lanzillotti-kimura_resonant_2009} and is compatible with the integration with quantum dots and quantum wells. However, the concepts presented in this work can be easily extended to other materials platforms.

Moreover, we demonstrated that multiple topological acoustic interface states can be generated simultaneously at different bandgap orders. Importantly, the generated interface states are robust against disorder in the unit cell thickness ratio across a broad frequency range that does not affect the associated Zak phases. A promising extension of this system might thus arise by introducing spatial periodicity to the system. Then, the platform developed here could potentially enable the development and study of synthetic dimensions, where high-order interface states would serve as an additional lattice dimension for the topological system~\cite{del_pino_non-hermitian_2022,mancini_observation_2015}. Furthermore, we demonstrated the presence of interface states in hybrid structures, i.e., structures combining two superlattices with bandgaps of different orders centered at the same frequency. The interface states in high-order bandgaps presented here can potentially be a useful tool to explore a full class of hybrid topological resonators. One could, for instance, exploit the versatility of hybrid topological resonators to generate multiple interface states at higher-order frequency-matched bandgaps, that are difficult to access by electronics or optics due to their respective dispersion relations. Overall, our results constitute an important step in the development of nanophononics for robust noise-insensitive communication, data processing, and quantum technologies.

\section{Acknowledgments}
The authors gratefully acknowledge M. Esmann for fruitful discussions and  support at an early stage of the project.  The authors acknowledge funding from European Research Council Consolidator Grant No.101045089 (T-Recs). This work was supported by the European Commission in the form of the H2020 FET Proactive project No. 824140 (TOCHA).  

\bibliography{MAIN}

\end{document}